\documentclass[conference]{IEEEtran}
\IEEEoverridecommandlockouts
\usepackage{cite}
\usepackage{amsmath,amssymb,amsfonts}
\usepackage{algorithmic}

\usepackage{graphicx}

\usepackage{textcomp}
\usepackage{xcolor}
\def\BibTeX{{\rm B\kern-.05em{\sc i\kern-.025em b}\kern-.08em
    T\kern-.1667em\lower.7ex\hbox{E}\kern-.125emX}}
\usepackage{amsthm}

\usepackage[colorlinks]{hyperref}
\usepackage[nameinlink]{cleveref}

\usepackage{pgfplots}
\usetikzlibrary{patterns}
\usetikzlibrary{external}

\newtheorem{theorem}{Theorem}
\newtheorem{lemma}{Lemma}
\newtheorem{corollary}{Corollary}
\newtheorem{remark}{Remark}

\def\BibTeX{{\rm B\kern-.05em{\sc i\kern-.025em b}\kern-.08em
    T\kern-.1667em\lower.7ex\hbox{E}\kern-.125emX}}
\begin{document}

\title{On Missing Mass Variance
\thanks{Supported by the FNR grant C17/IS/11613923}
}

\author{\IEEEauthorblockN{Maciej Skorski}
\IEEEauthorblockA{\textit{University of Luxembourg}
}
}

\maketitle

\begin{abstract}
The missing mass refers to the probability of elements not observed in a sample, and since the work of Good and Turing during WWII, has been studied extensively in many areas including ecology, linguistic, networks and information theory.

This work determines what is the \emph{maximal variance of the missing mass}, for any sample and alphabet sizes. The result helps in understanding the missing mass concentration properties.
\end{abstract}


\section{Introduction}

\subsection{Background}

Let $X_1,\ldots,X_n\sim^{IID} p$ be a sample from a distribution $(p_s)_{s
\in S}$ on a countable alphabet $S$, and $f_s = \#\{i: X_i = s\}$ be the empirical (observed) frequencies. The missing mass
\begin{align}
    M_0 = \sum_{s}p_s\mathbb{I}(f_s=0),
\end{align}
which quantifies how much of the population does not appear in the sample, is of interest to statistics~\cite{robbins1968estimating} and applied disciplines, such as: 
ecology~\cite{shen2003predicting,chao2003nonparametric,chao2013entropy,chao2017seen}, linguistic~\cite{efron1976estimating,mcneil1973estimating,thisted1987did,gale1995good}, archaeology~\cite{myrberg2015tale}, network design~\cite{budianu2003estimation,budianu2004good}, information theory~\cite{vu2007coverage,zhang2012entropy}, microbiology~\cite{hughes2001counting}
and bio-molecular modeling~\cite{mao2002poisson,koukos2014application}. The 
 problem was studied first by Good and Turing~\cite{good1953population}.

This work studies the maximal variance of the missing mass
\begin{align}\label{eq:mse_def}
    \max_{(p_s)\in\mathbb{P}(S)} \mathbf{Var}[M_0] = ?
\end{align}
given the alphabet size (equivalently, the constrained support):
\begin{align}
\# S = m,
\end{align}
where we allow for $m=+\infty$ and denote by $\mathbb{P}(S)$ the set of probability measures on $S$.
We are interested in deriving a formula valid for any sample size $n$ and any alphabet size $m$, with the relative error of at most $o(1)$ when $n$ grows.

The motivation is two-fold. First, our work continues the line of research on extreme missing mass properties \cite{rajaraman2017minimax,minimaxa2018improved,berend2012missing,berend2017expected}. Second, more importantly, understanding the variance extremes determines no-go regimes for concentration inequalities, which have been actively studied \cite{mcallester2000convergence,mcallester2003concentration,berend2013concentration,ben2017concentration,chandra2019concentration,skorskisub,skorski2020missing}; the need for such no-go results was highlighted in prior works~\cite{rajaraman2017minimax,minimaxa2018improved}. %The third is the focus on the alphabet size. %Setups from prior works on extreme missing mass properties

\subsection{Related Work}

No prior work has addressed the problem of finding \emph{sharp variance bounds} for the missing mass, even under no alphabet constraints. Below we review weaker results that are available.

The maximal variance is $\Theta(n^{-1})$ under no constraints ($m=+\infty$). This has been noticed in the line of work on concentration properties~\cite{mcallester2000convergence}. Furthermore, this asymptotic holds up to a constant for many distributions with certain regularity properties~\cite{ben2017concentration}.
The asymptotic $\Theta(n^{-1})$, when $m=+\infty$, can be also deduced from works that studied the Good-Turing estimator~\cite{rajaraman2017minimax,minimaxa2018improved}; unfortunately the estimator variance differs from our variance already by $\Omega(n^{-1})$, and so it is insufficient to give the sharp asymptotic. 

As we will see, in the context of our problem techniques from prior works are not sufficient. While certain tricks~\cite{rajaraman2017minimax,minimaxa2018improved} can used to handle the smaller order term, determining the leading asymptotic term involves subtle non-linear programming, mainly due to the support constraint.
Analogues programs from prior works on extreme missing mass properties luckily had simpler objectives and boundaries. For example, the problem~\cite{berend2012missing} is straightforward due to Schur-convexity~\cite{xia2009schur}; in~\cite{minimaxa2018improved} the reduced program involves only one equality constraint; in~\cite{rajaraman2017minimax} the constant was found up to a small gap, but with elementary methods avoiding optimization.

The reader interested in missing mass beyond the scope of our problem may also refer to other works~\cite{esty1982confidence,zhang2009asymptotic,mossel2015impossibility,cohen2021non,orlitsky2003always,orlitsky2015competitive,cohen2017cardinality,ayed2019good}.

\subsection{Summary of Contribution}
The contribution of this work can be summarized as follows:
\begin{itemize}
    \item The formula accurately approximating  missing mass variance approximation, with the poissonized version which is better suited for analysis and numerical evaluation.
    \item The characterization of the worst-case variance value, along with the maximizing distribution, for any sample and alphabet size. The maximal variance behaves like $\Theta(\min\{\frac{1}{m},\frac{1}{n}\})$, with the explicit leading constant. The maximizer is a mixture of a uniform distribution and a point mass, with the explicitly given proportions.
    \item An application of the extreme variance values to benchmarking results on missing mass concentration.
\end{itemize}
%~\cite{berend2012missing,rajaraman2017minimax,minimaxa2018improved} luckily had simpler objectives and feasibility sets, so that 

\section{Results}

\subsection{Accurate Variance Estimation}

We present the accurate bound on the missing mass variance, valid for any discrete distribution. The derivation is similar as for the formula for the Good-Turing estimator, obtained in \cite{rajaraman2017minimax}, although quantitatively quite different.
%we note that reusing that estimate would incurr to the gap of %$O(n^{-1})$.

\begin{theorem}[Missing Mass Variance]\label{thm:poisson}
We have:
\begin{multline}
\mathbf{Var}[M_0] =-n\left(\sum_{s}p_s^2(1-p_{s})^{n}\right)^2   + n\sum_s p_s^3(1-p_s)^{n} \\ + O(n^{-2}).
\end{multline}
\end{theorem}
We can further approximate the variance, similarly as for moments of occupancy numbers~\cite{gnedin2007notes,barbour2009small}, with "poissonized" terms. This makes the formula easier to 
analyze and also more stable numerically (via the "log-sum-exp" trick~\cite{blanchard2019accurate}). 
\begin{corollary}[Missing Mass Variance, Poissonized]\label{cor:poisson}
We have:
\begin{multline}
\mathbf{Var}[M_0] =-n\left(\sum_{s}p_s^2\mathrm{e}^{-np_s}\right)^2   + n\sum_s p_s^3\mathrm{e}^{-np_s} \\ + O(n^{-2}).
\end{multline}
\end{corollary}

\subsection{Extreme Variance Behavior}

Building on \Cref{cor:poisson}, we use non-linear optimization to characterize the maximal variance. The result is stated below:
\begin{theorem}\label{thm:main_opt}
Denote $m = \# S$, $b=\frac{m}{n}$. Consider the program: 
\begin{align} 
\begin{aligned}
\max_{w,c} && \alpha(w,c)=
    - w^2 c^2\mathrm{e}^{-2c} + w c^2\mathrm{e}^{-c}  \\
    \mathrm{s.t.} && 
    \begin{cases}
 0\leqslant w  \leqslant 1\\
    w \leqslant b c,
\end{cases}
\end{aligned}
\end{align}
and let $\alpha$ be its optimal value achieved at $w,c$. Then:
\begin{align}
     \max_{(p_s)\in\mathbb{P}(S)} \mathbf{Var}[M_0] = \frac{\alpha}{n} + O(n^{-2}),
\end{align}
with the maximum achieved at $(p_s)$ being the mixture of a uniform distribution on $\min\{\lceil wn/ c \rceil,m-1\}$ elements and a Dirac mass,
with the proportion $w$ and $1-w$ respectively.

For $m=+\infty$, the constraints reduce to $0\leqslant w\leqslant 1,0\leqslant c$.
\end{theorem}

\begin{figure}[h]
\begin{tikzpicture}[scale=0.8]
\begin{axis}[
    %title=,
    %hide axis,
    % 3d box=complete,
    axis lines = left,
    colormap/cool,
    xlabel = $w$,
    ylabel = $c$,
    y dir = reverse,
    z dir = reverse,
    zmax = 0.50,
    view={60}{-140},
    legend style={at={(0.0,0.9)},anchor=north west},
    enlargelimits = true,
    colorbar
]
\addplot3[
    surf,
    samples=50,
    domain=0:1,
    domain y = 0:5
]
{ -x^2*y^2*exp(-2*y) + x*y^2*exp(-y) };
\addlegendentry{$-w^2c^2\mathrm{e}^{-2c}+wc^2\mathrm{e}^{-c}$}
    \addplot3[
    black,ultra thick,dotted,
    domain=0:5,
    samples = 50,
    samples y = 0
]
    (
    {min(0.8*x,1)},
    {x},
    {-min(0.8*x,1)^2*x^2*exp(-2*x) + min(0.8*x,1)*x^2*exp(-x)}
    );
\end{axis}
\end{tikzpicture}
\caption{The optimization landscape of the problem in \Cref{thm:main_opt}.\\ 
The boundary of the feasible region is shown with the dotted line.}
\label{fig:landscape}
\end{figure}
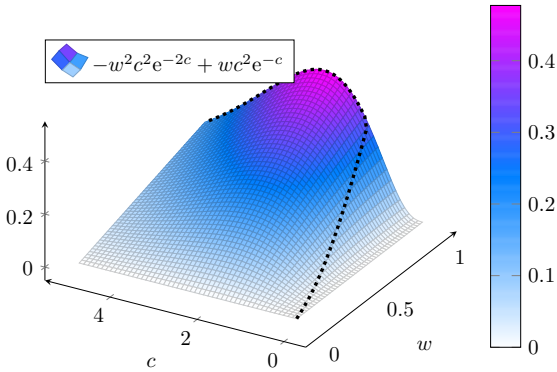

The shape of the worst-case distribution, a Uniform-Dirac mixture, appears
often for extreme problems in nonparameteric statistics~\cite{bu2018estimation} and could be expected; the proof is however quite subtle. The optimization is illustrated in \Cref{fig:landscape}.

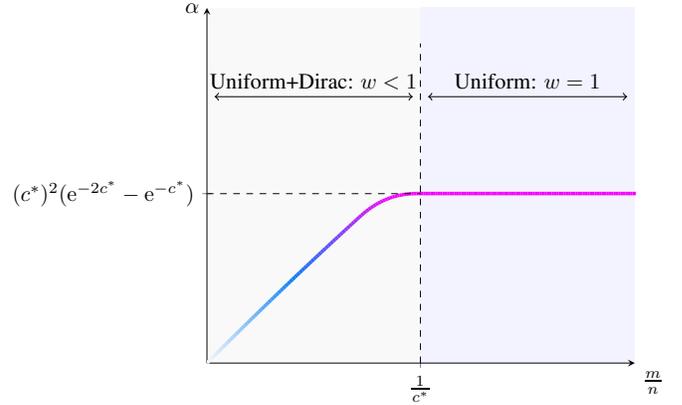
\begin{figure}[h]
\begin{tikzpicture}[scale=0.83]
\begin{axis}[
enlargelimits = true,
axis lines =middle,
xlabel = {$\frac{m}{n}$},
ylabel = {$\alpha$},
xtick={0, 0.449, 2},
xticklabels={0,$\frac{1}{c^{*}}$},
yticklabels={0,$(c^{*})^2(\mathrm{e}^{-2c^{*}}-\mathrm{e}^{-c^{*}})$},
ytick={0, 0.477},
ymax = 1,
xmax = 0.9,
colormap/cool,
every axis x label/.style={
    at={(ticklabel* cs:1.0)},
    anchor=north west,
},
every axis y label/.style={
    at={(ticklabel* cs:1.0)},
    anchor=east,
},
]
\addplot[mesh,ultra thick] table [x=b,y=val,col sep=comma] {miss_mass_var.csv};
\addplot[mark=none,dashed,black] coordinates {(0,0.477) (0.449,0.477)};
\addplot[mark=none,dashed,black] coordinates {(0.449,0) (0.449,0.9)};
\path[fill=gray,fill opacity=0.05] (axis cs:0,0) -- (axis cs:0,1) -- (axis cs:0.449,1)--(axis cs:0.449,0);
\path[fill=blue,fill opacity=0.05] (axis cs:0.449,0) -- (axis cs:0.449,1) -- (axis cs:0.9,1.72)--(axis cs:0.9,0);
%\node at (axis cs:0.5,0.75) {Uniform+Dirac};
%\node at (axis cs:1.5,0.75) {Uniform};
\node (n1) at (axis cs: 0,0.75) {};
\node (n2) at (axis cs: 0.449,0.75) {};
\node (n3) at (axis cs: 0.9,0.75) {};
\draw[<->] (n1)--(n2) node[midway,above] {Uniform+Dirac: $w<1$};
\draw[<->] (n2)--(n3) node[midway,above] {Uniform: $w=1$};
\end{axis}
\end{tikzpicture}
\caption{The constant $\alpha$ in the asymptotic $\Theta(n^{-1})$ for the maximal missing mass variance \eqref{eq:mse_def}, depending on the alphabet size $m=\#S$ (\Cref{cor:phase}).
The phase transition (the maximizer becomes uniform) occurs at $m = \frac{n}{c^{*}}$.}
\label{fig:phase}
\end{figure}

Finally, we give a reduction to a one-dimensional problem, which can be solved \emph{numerically} for any given $m,n$.
\begin{corollary}\label{cor:phase}
Let $c^{*}=2.26281...$ be the root of $2 - 2 e^c + c (-2 + e^c)=0$.
Then the optimal solution for \Cref{thm:main_opt} is:
\begin{align}
\alpha=
    \begin{cases}
    \max_{c\geqslant 0} -c^2\mathrm{e}^{-2c}+c^2\mathrm{e}^{-c} = 0.477\ldots & \frac{m}{n}\geqslant \frac{1}{c^{*}}\\
    \max_{0<c\leqslant \frac{1}{b}} -b^2 c^4 \mathrm{e}^{-2c}+b c^3\mathrm{e}^{-c} & \frac{m}{n} < \frac{1}{c^{*}}
    \end{cases},
\end{align}
and $w$ equals $1$ and respectively $b c$, where $c$ is the maximizer.
\end{corollary}
We see that the constant in the maximal variance asymptotic depends on the ratio of the sample size and alphabet size. Furthermore, for sufficiently large $n$ the Dirac part vanishes (phase transition), as illustrated in
\Cref{fig:phase}.

\subsection{Application: Gap in Missing Mass Concentration}

We show how our result can be used to benchmark concentration bounds. This solves the problem posed in~\cite{minimaxa2018improved}.

The state-of-the-art bounds for the concentration of $M_0$~\cite{ben2017concentration} give the following:
$M_0$ is sub-gamma with variance-factor $v=\Theta(n^{-1})$ and scale $c=O(n^{-1})$, and thus satisfies
the concentration bound $\Pr[|M_0-\mathbb{E}[M_0]|>\epsilon]\leqslant 2\exp\left(\frac{-\epsilon^2}{2v^2+2c\epsilon}\right)$ (in particular $2\mathrm{e}^{-\Omega(n\epsilon^2)}$ as we can assume $\epsilon<1$). However we know that $v^2 \geqslant \mathbf{Var}[M_0]$ for the sub-gamma bound, and this is sharp~\cite{boucheron2013concentration}. Using our results, we observe that:
\begin{itemize}
    \item The biggest variance is $\mathbf{Var}[M_0]\approx \frac{0.477}{n}$. In turn, by the numeric formula~\cite{boucheron2013concentration}, we have $v=\sum_s p_s^2(1-p_s)^{n} + n^{-1}\sum_s p_s(1-p_s)^{n}$ which can be as big as $v\approx \frac{0.839}{n}$. This leaves the gap of $\frac{0.362}{n}$ w.r.t. the "ideal" case.
    \item Missing mass is a sum of negatively-dependent variables, and the all prior works on missing mass concentration handle this by the IID majorization. However ignoring correlation terms gives the variance-term of at least $v\approx \sum_s p_s^2 ((1-p_s)^n - (1-p_s)^{2n})$, which can be as far as by $\Omega(n^{-1})$  from the "ideal" value of $\mathbf{Var}[M_0]$.
\end{itemize}

\section{Preliminaries}

We will need the mean-value theorem to obtain asymptotic bounds, as well as to localize and count roots of equations.
\begin{lemma}[Mean-Value Theorem~\cite{feng2013mean}]\label{lemma:mean_val_thm}
If a function $f(u)$ is continuous on the closed interval $[a,b]$ and differentiable on the open interval $(a,b)$, then there exists a point $c\in(a,b)$ such that $\frac{\partial f}{\partial u}(c) = \frac{f(b)-f(a)}{b-a}$.
\end{lemma}
We will also need the facts below in our estimations:
\begin{lemma}[Generalized-Bernoulli Inequality]\label{lemma:binomial_series}
Let $1\leqslant k\leqslant n$ be integers. Then for any $-1\leqslant u\leqslant 0$ it holds that:
$1+nu \leqslant (1+u)^n \leqslant 1+nu+O(nu^2)$.
\end{lemma}

\begin{lemma}[Taylor's Expansion with Remainder]\label{lemma:taylor_remainder}
Let $f$ be a twice differentiable real function, then $f(x) = f(x_0) + \frac{\partial f}{\partial x}(x_0)\cdot (x-x_0) + \frac{ \frac{\partial^2 f}{\partial x^2}(z)}{2}\cdot (x-x_0)^2$ for some $z\in [x,x_0]$.
\end{lemma}

Another fact handles the commonly occurring expressions:
\begin{lemma}[Occupancy-like Expressions]\label{lemma:exp_sums}
Let $k,n\geqslant 1$. Then $p^k(1-p)^n$ for $p\in [0,1]$ is maximized for $p=\frac{k}{k+n}$, 
and $p^k\mathrm{e}^{-np}$ for $p\in [0,1]$ is maximized for $p=\frac{k}{n}$. In particular if $(p_s)$ is a probability distribution, then $\sum_s p_s^k(1-p_s)^{n} = O(n^{1-k})$ and $\sum_s p_s^k\mathrm{e}^{-np_s} = O(n^{1-k})$ when $k=O(1)$.
\end{lemma}
Finally, we will need some theory of non-linear programming. For a detailed discussion we refer to books~\cite{boyd2004convex,biegler2010nonlinear,bazaraa2013nonlinear}.
\begin{lemma}[KKT Conditions]\label{lemma:KKT}
Consider the program
\begin{align}
\begin{aligned}
\max && f(x) \\
\mathrm{s.t.} &&
\begin{cases}
h_i(x) = 0,& i\in I\\
g_j(x)\leqslant 0,& i\in J
\end{cases}
\end{aligned}
\end{align}
with differentiable real functions $f$,$(h_i)_{i\in I},(g_j)_{j\in J}$ in variables $x=x_1,\ldots,x_d$.
If the maximum occurs at $x$, then:
\begin{align}
\frac{\partial}{\partial x}f(x) = \sum_{i} \lambda_i \frac{\partial}{\partial x}h_i(x) +\sum_{j} \mu_j \frac{\partial}{\partial x}g_j(x),
\end{align}
where $\frac{\partial}{\partial x} = (\frac{\partial}{\partial x_1}\ldots\frac{\partial}{\partial x_d})$,
for $\lambda_i\in\mathbb{R}$, $\mu_j\geqslant 0$ such that
\begin{align}
\lambda_i \in\mathbb{R},\ \mu_j\geqslant 0,\ \mu_j g_j(x) = 0,
\end{align}
provided that regularity conditions hold at $x$.
\end{lemma}
We briefly remind the optimization terminology. The function $f$ is called objective, and any $x\in\mathbb{R}^d$ satisfying the constraints is called feasible. The optimal value is also called the program value. The constraint $h_i$ respectively $g_j$ is called active at $x$, when $h_i(x)=0$, respectively $g_j(x)=0$.

\begin{remark}[Constraints Qualification]\label{rem:licq}
The KKT conditions hold for optimal $x$ when the active constraints gradients are linearly independent (LICQ) or for affine constraints (LQC).
\end{remark}

\begin{remark}[Extreme Values~\cite{lovric2007vector}]\label{lemma:extreme_val_theorem}
A continuous function on a compact subset of $\mathbb{R}^d$ achieves its maximum and minimum.
\end{remark}

\section{Proofs}

\subsection{Proof of \Cref{thm:poisson}}

Let $\xi_s$ be the indicator that $s$ does not occur in the sample.
By \Cref{lemma:binomial_series}:
\begin{align}
\begin{split}
    \mathbf{Var}[\xi_s] &= (1-p_s)^n-((1-p_s)^{2})^n \\
    & = (1-p_s)^n(1-(1-p_s)^n) \\
    & = n p_s(1-p_s)^n (1+O(n p_s)).
\end{split}
\end{align}
In turn, for $s\not=s'$ by 
\Cref{lemma:mean_val_thm}:
\begin{align}
\begin{split}
\mathbf{Cov}[\xi_s,\xi_{s'}]&=(1-p_s-p_{s'})^n-(1-p_s)(1-p_{s'})^n \\
& = -np_s p_{s'}(1-p_s)^n(1-p_{s'})^{n-1} + \\
& \quad + O\left(n^2p_s^2p_{s'}^2(1-p_s)^{n-2}(1-p_{s'})^{n-2}\right).
\end{split}
\end{align}
These bounds imply that:
\begin{align}
    \sum_s p_s^2\mathbf{Var}[\xi_s] = n \sum_s p_s^3(1-p_s)^n + O(n^{-2}),
\end{align}
where we used \Cref{lemma:exp_sums} to estimate the $O(\cdot)$ term, and
\begin{multline}
    \sum_{s\not=s'}p_s p_{s'}\mathbf{Cov}[\xi_s,\xi_{s'}] = \\
    -n\sum_{s\not=s'}p_s^2p_{s'}^2(1-p_s)^{n-2}(1-p_{s'})^{n-2} + O(n^{-2}),
\end{multline}
where we estimated the $O(\cdot)$ term using
$n^2\sum_{s\not=s'}p_s^3p_{s'}^3(1-p_s)^{n-2}(1-p_{s'})^{n-2}<(n\sum_s p_s^3(1-p_s)^{n-2})^2=O(n^{-2})$ due to \Cref{lemma:exp_sums}; since
$n\sum_s p_s^4(1-p_s)^{2(n-2)}=O(n^{-2})$ by \Cref{lemma:exp_sums}, we obtain the following simpler expression:
\begin{align}
    \sum_{s\not=s'}p_s p_{s'}\mathbf{Cov}[\xi_s,\xi_{s'}] = -n(\sum_s p_s^2(1-p_s)^{n-2})^2 + O(n^{-2}).
\end{align}
Since $\sum_s p_s^2(1-p_s)^{n-2} = \sum_s p_s^2(1-p_s)^n + O(n^{-2})$, as seen by $p_s^2(1-p_s)^n = p_s^2(1-p_s)^{n-2}(1-2p_s+O(p_s^2))$ and \Cref{lemma:exp_sums}, we can replace the exponent $n-2$ with $n$.

By the identity $\mathbf{Var}[M] = \sum_s p_s\mathbf{Var}[\xi_s] + \sum_{s\not=s'}p_s p_{s'}\mathbf{Cov}[\xi_s,\xi_{s'}]$, and the bounds we derived above for the two pars of the summation:
\begin{multline}
\mathbf{Var}[M_0] = -n\left(\sum_{s}p_s^2(1-p_{s})^{n}\right)^2   + n\sum_s p_s^3(1-p_s)^{n} \\ + O(n^{-2}).
\end{multline}

At the final step we observe $|\mathrm{e}^{-n p}-(1-p)^n |\leqslant O(n p^2\mathrm{e}^{-np})$ by \Cref{lemma:mean_val_thm} applied to $f(u)=u^n$. By \Cref{lemma:exp_sums}, we conclude that in the above identity we can replace 
$1-p$ by $\mathrm{e}^{-p}$ making the error of $O(n\sum_s p_s^4\mathrm{e}^{-np_s})=O(n^{-2})$ in the first sum
and $O(n\sum_s p_s^5\mathrm{e}^{-np_s})=O(n^{-3})$ in the third sum. Both sums are $O(n^{-1})$, thus the total additive error is $O(n^{-2})$.

\subsection{Proof of \Cref{thm:main_opt}}

\subsubsection{Non-linear Programming Formulation}
In view of \Cref{cor:poisson}, we need to solve the optimization program:
\begin{align}\label{eq:main_program}
\begin{aligned}
\max_{(p_s)} &&
    -n\left(\sum_s p_s^2\mathrm{e}^{-n p_s}\right)^2 +  n\sum_s p_s^3\mathrm{e}^{-np_s} \\
    \textrm{s.t.} && \sum_s p_s = 1,\ p_s\geqslant 0.
\end{aligned}
\end{align}

Observe that $(p_s)$ may be not finite-dimensional, but rather countable,
so it is not clear whether the maximum is achieved on a feasible point. However, the objective is bounded, so the supremum may be achieved on a sequence of feasible points.

\subsubsection{Maximizer is 4-Mixture}

We will show that every feasible $(p_s)$ with at least 4 distinct non-zero values can be improved, that is replaced by a feasible $(p_s)$ with the strictly bigger objective value. To this end we fix the values of $p_s$ on all indices except a finite subset $S'$ and consider the program
\begin{align}
\begin{aligned}
\max &&
    -n\left(\sum_{s\in S'} p_s^2\mathrm{e}^{-n p_s}+B\right)^2 +  n\sum_{s\in S'} p_s^3\mathrm{e}^{-np_s} + C \\
    \textrm{s.t.} && \sum_s p_s = A,\ p_s\geqslant 0.
\end{aligned}
\end{align}
for some $A,B,C>0$. It suffices to prove that regardless of values of $A,B,C$, the maximum is achieved for $(p_s)_{s\in S'}$ which takes at most 3 distinct non-zero values.  By \Cref{lemma:extreme_val_theorem} the maximum is indeed achieved at some feasible point. By \Cref{lemma:KKT} the \emph{non-zero components} of the optimal point $(p_s)_{s'}$ are solutions of the following equation in $u$:
\begin{align}
    -2n \cdot D\cdot \frac{\partial}{\partial u}[u^2\mathrm{e}^{-nu}] + n
    \cdot \frac{\partial}{\partial u}[u^3 \mathrm{e}^{-nu}]  = \lambda,
\end{align}
where $D = 2(\sum_{s\in S'}p_s^2\mathrm{e}^{-np_s}+B)$, for some constant $\lambda$. 
The equation is of the form 
$P(u)\mathrm{e}^{-n u} = \lambda$ where $P(u)$ is a polynomial of degree 3. 
We claim that such an equation has at most 4 solutions; to this end it suffices to show that the derivative of $P(u)\mathrm{e}^{-nu}$ has at most 3 solutions; indeed by \Cref{lemma:mean_val_thm} between two points with equal function value there is a root of its derivative (a.k.a. Rolle's Theorem). But the derivative is $P'(u)\mathrm{e}^{-nu}$ for a degree-3 polynomial $P'$, and has at most 3 roots by the Fundamental Theorem of Algebra.

This claim shows that, the equivalent program is
\begin{align}\label{eq:main_program_compact}
\begin{aligned}
\max_{(p_s),(k_s)} &&
    -n\left(\sum_{s=1}^{4} k_s p_s^2\mathrm{e}^{-n p_s}\right)^2 +  n\sum_{s=1}^{4} k_s p_s^3\mathrm{e}^{-np_s}  \\
    \textrm{s.t.} && 
    \begin{cases}
    \sum_{s=1}^{4} k_s p_s = 1\\
    p_s\geqslant 0 \\
    k_s\geqslant 0,\ k_s \in\mathbb{Z},
\end{cases}
\end{aligned}
\end{align}
whether the maximum in \eqref{eq:main_program} is achieved at a concrete point, or in the limit on a sequence of points. We can state it equivalently: when seeking the maximizer we can restrict the feasible set to mixtures of (at most) 4 uniform distributions.

\subsubsection{Relaxation}

We prove, as before, that \eqref{eq:main_program_compact} holds when $S$ is bounded. Let $m=\#S$, consider the relaxed program:
\begin{align}\label{eq:main2_program_compact_relaxed}
\begin{aligned}
\max_{(p_s),(k_s)} &&
    -n\left(\sum_{s=1}^{4} k_s p_s^2\mathrm{e}^{-n p_s}\right)^2 +  n\sum_{s=1}^{4} k_s p_s^3\mathrm{e}^{-np_s}  \\
    \textrm{s.t.} && 
    \begin{cases}
    \sum_{s=1}^{4} k_s p_s \leqslant 1\\
    \sum_{s=1}^{4}k_s \leqslant m \\
    p_s\geqslant 0 \\
    k_s\geqslant 0.
\end{cases}
\end{aligned}
\end{align}
We prove that the global maximum is achieved. Consider the supremum of the objective under the constraints; it can be approached in the limit of the objective values on the sequence of feasible points $(k^j_s),(p^j_s)$ for $j=1,2,\ldots$. By the diagonal argument (passing to a subequence) we can assume that for every $s$ the sequences $k^j_s,p^j_s$ are monotone.
Observe now that $p^j_s\to \infty$ for $j\to\infty$, then the objective in the limit and the feasibility does not change when we replace $p^j_s = 0$; this is because 
$p^2\mathrm{e}^{-np},p^3\mathrm{e}^{-np}\to 0$ for $p\to\infty$. We can thus assume that $(p^j_s)$ are bounded. Suppose now that $k^j_s\to \infty$ for some $s$, when $j\to\infty$. Then it must be that $p^j_s\to 0$ because $k_sp_s\leqslant 1$; then the objective in the limit and the feasibility do not change when we replace $k^j_s = 0$ for all $j$. We can thus assume that $k_s^j$ are also bounded. Since $p^j_s$ and $k^j_s$ are monotone and bounded, they converge to finite values $p^{*}_s,k^{*}_s$. By the continuity, the sequence of objective values converges to the objective on $p^{*}_s,k^{*}_s$.
Furthermore, these values are feasible, as the constraint set is closed. This shows that the supremum is achieved.

\subsubsection{Solving Relaxation}

Note that the constraints are bi-linear; thus we can apply the KKT conditions
for with respect to $p_s$ with optimal $k_s$, and with respect to $k_s$ with fixed optimal $p_s$. At the optimal solution $(p_s),(k_s)$ for $s$ such that $p_s k_s > 0$ we have that $u=p_s$ solves the system of equations
\begin{align}\label{eq:system2}
\begin{aligned}
  g(u) &= \lambda u + \mu \\
    \frac{\partial g}{\partial u}(u) &= \lambda 
\end{aligned},
\end{align}
where we introduced the auxiliary function
\begin{align}
    g(u)\triangleq (u^3-A u^2)\mathrm{e}^{-n u}
\end{align}
with $A = 2\sum_s k_s p_s^2\mathrm{e}^{-np_s}$. The first equation in \eqref{eq:system2} is the KKT condition for $(k_s)$ divided by $n$, and the second equation in \eqref{eq:system2} is the KKT condition for $(p_s)$ divided by $n$. Geometrically, \eqref{eq:system2} means that the line $u\to \lambda u + \mu$ is tangent to the plot of $g$ at $u$. 
We will now argue that this does not happen for \emph{distinct} values of $u$, so the system has one solution.

\begin{figure}
\resizebox{0.7\linewidth}{!}{
\begin{tikzpicture}
\begin{axis}[
    axis lines*=left,
    xlabel = $u$,
    ylabel = {$g(u)$},
    %ymax = 0.01,
    %legend pos=south east
    yticklabel style={/pgf/number format/fixed}
]
%Below the red parabola is defined
\addplot [
    domain= 0:1.5, 
    samples=100, 
    color=red,
]
{(x^3-0.2*x^2)*exp(-10*x)};
\addlegendentry{$(u^3-Au^2)\mathrm{e}^{-n u}$}
\end{axis}
\end{tikzpicture}
}
\caption{The auxliary function $g(u)=(u^3-Au^2)\mathrm{e}^{-nu}$, for $A\in (0,1)$ and integer $n>1$. In this example $n=10$ and $A=0.2$.}
\label{fig:function2}
\end{figure}
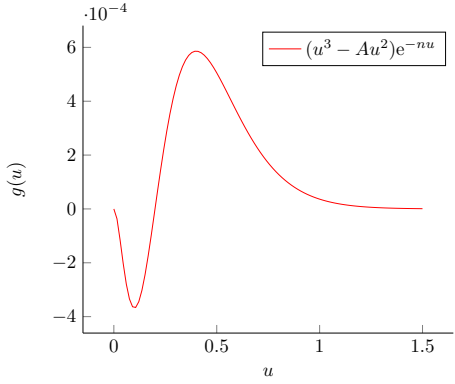

This is best seen on \Cref{fig:function2}, but the formal argument is rather subtle. Suppose that \eqref{eq:system2} has two solutions $u_1<u_2$. 
The derivative  $\frac{\partial g}{\partial u}(u)=-\mathrm{e}^{-n u} u (A (2 - n u) + u (-3 + n u))$ has 3 roots which split the domain into 
intervals $I_1,I_2,I_3$ such that $g$ is decreasing on $I_1$, increasing on $I_2$ and decreasing on $I_3$; the endpoints are the roots 
$0$, $\frac{3 + A n - \sqrt{9 - 2 A n + A^2 n^2}}{2 n}$, $\frac{3 + A n + \sqrt{9 - 2 A n + A^2 n^2}}{2 n}$. We then claim that on every interval $I_j$
the derivative takes every value at most twice. Otherwise, one of the intervals contains two roots of the second derivative
$\frac{\partial^2 g}{\partial u^2}$, by \Cref{lemma:mean_val_thm} (applied to the repeating values); since every interval contains at least one root of the second derivative, by \Cref{lemma:mean_val_thm} applied to the endpoints (for $I_3$, we use $+\infty$ as the right endpoint), in total this gives 4 roots.
On the other hand the roots of $\frac{\partial^2 g}{\partial u^2}$ satisfy the polynomial equation of order 3 with at most 3 roots, a contradiction. 

Thus, for the optimal solution of \eqref{eq:main2_program_compact_relaxed} 
$k_s p_s \not=0$ for only one $s$. In other words, the program is equivalent to:
\begin{align}\label{eq:main2_program_compact_relaxed_2d} 
\begin{aligned}
\max_{p,k} &&
    -n k^2 p^4\mathrm{e}^{-2n p} +  n k p^3\mathrm{e}^{-np}  \\
    \textrm{s.t.} && 
    \begin{cases}
 k  p  \leqslant 1\\
    0\leqslant k\leqslant  m \\
    0 \leqslant p.
\end{cases}
\end{aligned}
\end{align}

\subsubsection{Relaxation Gap}

Consider the optimal solution $(p,k)$ for \eqref{eq:main2_program_compact_relaxed_2d}. We will slightly modify it
so that it is feasible for \eqref{eq:main_program_compact}, but 
the relaxed objective does not change much. Define:
\begin{align}\label{eq:best_construction}
\begin{aligned}
    k_1 = \min\{\lceil k \rceil,m-1\},\ p_1 = p \\
    k_2=1,\ p_2 = 1-k_1p_1\\
    k_3,k_4,p_3,p_4 = 0.
\end{aligned}
\end{align}
Consider the following quantities:
\begin{align}
\begin{aligned}
    P' & = -n\left(\sum_s k_sp_s^2\mathrm{e}^{-np_s}\right)^2 + n\sum_s k_sp_s^3\mathrm{e}^{-np_s} \\
    P & = -n ( kp\mathrm{e}^{-np})^2 + n k p^3\mathrm{e}^{-np}
\end{aligned}.
\end{align}
We then have:
\begin{align}
\begin{aligned}
    \Delta_1 &\triangleq n(\sum_s k_sp_s^2\mathrm{e}^{-np_s})^2- n(k p^2 \mathrm{e}^{-np})^2  \\
    & \leqslant n(\sum_s k_sp_s^2\mathrm{e}^{-np_s})- k p^2 \mathrm{e}^{-np})\cdot O(n^{-1})\\
    &= ((k_1-k) p^2\mathrm{e}^{-np}  + O(n^{-2}))\cdot O(1) \\
    & \leqslant O(n^{-2}).
\end{aligned}
\end{align}
Where in the second line we use the identity $x^2-y^2=(x-y)(x+y)$
with $x=\sum_s k_s p_s^2\mathrm{e}^{-np_s}$, $y=kp^2\mathrm{e}^{-np}$
and the fact that $x+y=O(n^{-1})$ by \Cref{lemma:exp_sums}; in the third line we use again \Cref{lemma:exp_sums} and $|k_1-k|\leqslant O(1)$ (by the definition of $k_1$).

Similarly we obtain:
\begin{align}
\begin{aligned}
    \Delta_2 &\triangleq n k p^3\mathrm{e}^{-np} - n\sum_s k_s p_s^3\mathrm{e}^{-np_s} \\
    & = n (k-k_1) p^3\mathrm{e}^{-np} - O(n^{-2}) \\
    & \leqslant O(n^{-2}),
\end{aligned}
\end{align}
where we used \Cref{lemma:exp_sums} (twice) and $|k_1-k|\leqslant O(1)$.

Therefore $  P- P' = \Delta_1 + \Delta_2 \leqslant O(n^{-2})$, 
since also $ P' \leqslant P$ (by relaxation) we finally obtain:
\begin{align}
|P - P'| \leqslant O(n^{-2}).
\end{align}

\subsubsection{Concluding Result}

By the change of variables $w = kp$, $c = pn$ and $b\triangleq\frac{m}{n}$ we can rewrite \eqref{eq:main2_program_compact_relaxed_2d} as follows:
\begin{align}\label{eq:simple_2d}
\begin{aligned}
\max_{w,c} &&
    \frac{- w^2 c^2\mathrm{e}^{-2c} + w c^2\mathrm{e}^{-c}}{n}  \\
    \textrm{s.t.} && 
    \begin{cases}
 0\leqslant w  \leqslant 1\\
    w \leqslant b c.
\end{cases}
\end{aligned}
\end{align}
This proves the first part of \Cref{thm:main_opt} (the maximal value); note that the relaxation gap of $O(n^{-2})$ with respect to the program
\eqref{eq:main_program} and the difference of $O(n^{-2})$ between this program and the variance, give the total error of $O(n^{-2})$.

For the second part (the maximizing distribution) we use the construction \eqref{eq:best_construction}, which gives the stated characterization.

\subsection{Proof of \Cref{cor:phase}}

Consider the program \eqref{eq:simple_2d}. The maximum is achieved, 
as inherited from \eqref{eq:main2_program_compact_relaxed} (it follows also directly: the objective is non-negative and goes to zero when $c\to \infty$, uniformly in $w$); in fact is is achieved for $w,c>0$ (then the objective, under the constraints, is positive; for $wc=0$ it equals zero).

We first note that the optimal solution must be on the boundary. Otherwise if $0<w<1$ and $0<w<bc$ the first order conditions give
$w=\frac{\mathrm{e}^c}{2}$
and $\frac{1}{n} c \mathrm{e}^{-2 c} w (2 (c  -1)w -(c - 2) \mathrm{e}^c  )=0$, which leads to $w=\frac{\mathrm{e}^c}{2}$ and 
$2 (c  -1)w -(c - 2) \mathrm{e}^c )=0$, which in turn leads to a contradiction that $c-1=c-2$.

Thus our optimal value equals $\frac{1}{n}$ times the maximum of the following two cases which determine the constant $\alpha$: either
\begin{align}\label{case:1}
    \max_{c\geqslant \frac{1}{b}} -c^2\mathrm{e}^{-2c}+c^2\mathrm{e}^{-c}
\end{align}
or
\begin{align}\label{case:2}
    \max_{0<c\leqslant \frac{1}{b}} -b^2 c^4 \mathrm{e}^{-2c}+b c^3\mathrm{e}^{-c}.
\end{align}
For numerical optimization below we used the software \cite{2020SciPy-NMeth}.

Suppose that $b\leqslant \frac{1}{c^{*}}$. Let $g(c)=-c^2\mathrm{e}^{-2c}+c^2\mathrm{e}^{-c}$; this function has the unique maximizer $c^{*}\approx 2.26281$, is increasing for $0<c<c^{*}$ and decreasing for $c>c^{*}$. Then \eqref{case:1} equals 
$g(1/b)$ which matches the objective of \eqref{case:2} at $c=\frac{1}{b}$;
thus the optimal value of \eqref{case:1} is smaller or equal to that of \eqref{case:2}.

Suppose that $b>\frac{1}{c^{*}}$. Then \eqref{case:1} equals $g(c^{*})$.
We will prove that \eqref{case:2} is smaller or equal to $g(c^{*})$.
This is true when $\frac{\mathrm{e}^c}{2}<1$, because then
\eqref{case:2} is at most $c^2\mathrm{e}^{-c}$ maximized over $c\leqslant \log 2$, which is $ 0.24...$ (we used $bc\leqslant 1$). We can assume that $1\leqslant \frac{\mathrm{e}^c}{2}$;
note that the objective of \eqref{case:2} increases in $b$ when 
$b\leqslant \frac{\mathrm{e}^{c}}{2c}$, and under the constraint we have 
$0\leqslant b\leqslant \frac{1}{c} \leqslant \frac{\mathrm{e}^c}{2c}$. Thus,
\eqref{case:2} can be upper-bounded by setting $b=\frac{1}{c}$ in the objective and removing the constraint on $c$. This gives the upper bound of $\max_c g(c) =g(c^{*})$, and finishes the proof.% of \Cref{cor:phase}.

\section{Conclusion}

This work shows the \emph{exact} behavior of the missing mass variance in the extreme case. The result can be used to benchmark concentration results and to understand no-go regimes. The optimization techniques can be possibly generalized and used to obtain similar characterization for other occupancy numbers, such as small counts recently studied~\cite{battiston2021consistent}.

\newpage

\bibliographystyle{IEEEtran}
\bibliography{citations}

\begin{thebibliography}{10}
\providecommand{\url}[1]{#1}
\csname url@samestyle\endcsname
\providecommand{\newblock}{\relax}
\providecommand{\bibinfo}[2]{#2}
\providecommand{\BIBentrySTDinterwordspacing}{\spaceskip=0pt\relax}
\providecommand{\BIBentryALTinterwordstretchfactor}{4}
\providecommand{\BIBentryALTinterwordspacing}{\spaceskip=\fontdimen2\font plus
\BIBentryALTinterwordstretchfactor\fontdimen3\font minus
  \fontdimen4\font\relax}
\providecommand{\BIBforeignlanguage}[2]{{%
\expandafter\ifx\csname l@#1\endcsname\relax
\typeout{** WARNING: IEEEtran.bst: No hyphenation pattern has been}%
\typeout{** loaded for the language `#1'. Using the pattern for}%
\typeout{** the default language instead.}%
\else
\language=\csname l@#1\endcsname
\fi
#2}}
\providecommand{\BIBdecl}{\relax}
\BIBdecl

\bibitem{robbins1968estimating}
H.~E. Robbins \emph{et~al.}, ``Estimating the total probability of the
  unobserved outcomes of an experiment,'' \emph{The Annals of Mathematical
  Statistics}, vol.~39, no.~1, pp. 256--257, 1968.

\bibitem{shen2003predicting}
T.-J. Shen, A.~Chao, and C.-F. Lin, ``Predicting the number of new species in
  further taxonomic sampling,'' \emph{Ecology}, vol.~84, no.~3, pp. 798--804,
  2003.

\bibitem{chao2003nonparametric}
A.~Chao and T.-J. Shen, ``Nonparametric estimation of shannon’s index of
  diversity when there are unseen species in sample,'' \emph{Environmental and
  ecological statistics}, vol.~10, no.~4, pp. 429--443, 2003.

\bibitem{chao2013entropy}
A.~Chao, Y.~Wang, and L.~Jost, ``Entropy and the species accumulation curve: a
  novel entropy estimator via discovery rates of new species,'' \emph{Methods
  in Ecology and Evolution}, vol.~4, no.~11, pp. 1091--1100, 2013.

\bibitem{chao2017seen}
A.~Chao, R.~K. Colwell, C.-H. Chiu, and D.~Townsend, ``Seen once or more than
  once: Applying good--turing theory to estimate species richness using only
  unique observations and a species list,'' \emph{Methods in Ecology and
  Evolution}, vol.~8, no.~10, pp. 1221--1232, 2017.

\bibitem{efron1976estimating}
B.~Efron and R.~Thisted, ``Estimating the number of unseen species: How many
  words did shakespeare know?'' \emph{Biometrika}, vol.~63, no.~3, pp.
  435--447, 1976.

\bibitem{mcneil1973estimating}
D.~R. McNeil, ``Estimating an author's vocabulary,'' \emph{Journal of the
  American Statistical Association}, vol.~68, no. 341, pp. 92--96, 1973.

\bibitem{thisted1987did}
R.~Thisted and B.~Efron, ``Did shakespeare write a newly-discovered poem?''
  \emph{Biometrika}, vol.~74, no.~3, pp. 445--455, 1987.

\bibitem{gale1995good}
W.~A. Gale and G.~Sampson, ``{G}ood-{T}uring frequency estimation without
  tears,'' \emph{Journal of quantitative linguistics}, vol.~2, no.~3, pp.
  217--237, 1995.

\bibitem{myrberg2015tale}
N.~Myrberg~Burstr{\"o}m, ``A tale of buried treasure, some good estimations,
  and golden unicorns: The numismatic connections of alan turing.'' 2015.

\bibitem{budianu2003estimation}
C.~Budianu and L.~Tong, ``Estimation of the number of operating sensors in
  sensor network,'' in \emph{The Thrity-Seventh Asilomar Conference on Signals,
  Systems \& Computers, 2003}, vol.~2.\hskip 1em plus 0.5em minus 0.4em\relax
  IEEE, 2003, pp. 1728--1732.

\bibitem{budianu2004good}
------, ``{G}ood-{T}uring estimation of the number of operating sensors: a
  large deviations analysis,'' in \emph{2004 IEEE International Conference on
  Acoustics, Speech, and Signal Processing}, vol.~2.\hskip 1em plus 0.5em minus
  0.4em\relax IEEE, 2004, pp. ii--1029.

\bibitem{vu2007coverage}
V.~Q. Vu, B.~Yu, and R.~E. Kass, ``Coverage-adjusted entropy estimation,''
  \emph{Statistics in medicine}, vol.~26, no.~21, pp. 4039--4060, 2007.

\bibitem{zhang2012entropy}
Z.~Zhang, ``Entropy estimation in turing's perspective,'' \emph{Neural
  computation}, vol.~24, no.~5, pp. 1368--1389, 2012.

\bibitem{hughes2001counting}
J.~B. Hughes, J.~J. Hellmann, T.~H. Ricketts, and B.~J. Bohannan, ``Counting
  the uncountable: statistical approaches to estimating microbial diversity,''
  \emph{Applied and environmental microbiology}, vol.~67, no.~10, pp.
  4399--4406, 2001.

\bibitem{mao2002poisson}
C.~X. Mao and B.~G. Lindsay, ``A {P}oisson model for the coverage problem with
  a genomic application,'' \emph{Biometrika}, vol.~89, no.~3, pp. 669--682,
  2002.

\bibitem{koukos2014application}
P.~I. Koukos and N.~M. Glykos, ``On the application of {G}ood-{T}uring
  statistics to quantify convergence of biomolecular simulations,''
  \emph{Journal of chemical information and modeling}, vol.~54, no.~1, pp.
  209--217, 2014.

\bibitem{good1953population}
I.~J. Good, ``The population frequencies of species and the estimation of
  population parameters,'' \emph{Biometrika}, vol.~40, no. 3-4, pp. 237--264,
  1953.

\bibitem{rajaraman2017minimax}
N.~Rajaraman, A.~Thangaraj, and A.~T. Suresh, ``Minimax risk for missing mass
  estimation,'' in \emph{2017 IEEE International Symposium on Information
  Theory (ISIT)}.\hskip 1em plus 0.5em minus 0.4em\relax IEEE, 2017, pp.
  3025--3029.

\bibitem{minimaxa2018improved}
J.~Acharya, Y.~Bao, Y.~Kang, and Z.~Sun, ``Improved bounds for minimax risk of
  estimating missing mass,'' in \emph{2018 IEEE International Symposium on
  Information Theory (ISIT)}.\hskip 1em plus 0.5em minus 0.4em\relax IEEE,
  2018, pp. 326--330.

\bibitem{berend2012missing}
D.~Berend and A.~Kontorovich, ``The missing mass problem,'' \emph{Statistics \&
  Probability Letters}, vol.~82, no.~6, pp. 1102--1110, 2012.

\bibitem{berend2017expected}
D.~Berend, A.~Kontorovich, and G.~Zagdanski, ``The expected missing mass under
  an entropy constraint,'' \emph{Entropy}, vol.~19, no.~7, p. 315, 2017.

\bibitem{mcallester2000convergence}
D.~A. McAllester and R.~E. Schapire, ``On the convergence rate of
  {G}ood-{T}uring estimators.'' in \emph{COLT}, 2000, pp. 1--6.

\bibitem{mcallester2003concentration}
D.~McAllester and L.~Ortiz, ``Concentration inequalities for the missing mass
  and for histogram rule error,'' \emph{Journal of Machine Learning Research},
  vol.~4, no. Oct, pp. 895--911, 2003.

\bibitem{berend2013concentration}
D.~Berend, A.~Kontorovich \emph{et~al.}, ``On the concentration of the missing
  mass,'' \emph{Electronic Communications in Probability}, vol.~18, 2013.

\bibitem{ben2017concentration}
A.~Ben-Hamou, S.~Boucheron, M.~I. Ohannessian \emph{et~al.}, ``Concentration
  inequalities in the infinite urn scheme for occupancy counts and the missing
  mass, with applications,'' \emph{Bernoulli}, vol.~23, no.~1, pp. 249--287,
  2017.

\bibitem{chandra2019concentration}
P.~Chandra and A.~Thangaraj, ``Concentration and tail bounds for missing
  mass,'' in \emph{2019 IEEE International Symposium on Information Theory
  (ISIT)}.\hskip 1em plus 0.5em minus 0.4em\relax IEEE, 2019, pp. 1862--1866.

\bibitem{skorskisub}
M.~Skorski, ``On sub-gaussian concentration of missing mass.''

\bibitem{skorski2020missing}
------, ``Missing mass concentration for markov chains,'' \emph{arXiv preprint
  arXiv:2001.03603}, 2020.

\bibitem{xia2009schur}
W.-F. Xia and Y.-M. Chu, ``Schur-convexity for a class of symmetric functions
  and its applications,'' \emph{Journal of Inequalities and Applications}, vol.
  2009, pp. 1--15, 2009.

\bibitem{esty1982confidence}
W.~W. Esty, ``Confidence intervals for the coverage of low coverage samples,''
  \emph{The Annals of Statistics}, pp. 190--196, 1982.

\bibitem{zhang2009asymptotic}
C.-H. Zhang, Z.~Zhang \emph{et~al.}, ``Asymptotic normality of a nonparametric
  estimator of sample coverage,'' \emph{The Annals of Statistics}, vol.~37,
  no.~5A, pp. 2582--2595, 2009.

\bibitem{mossel2015impossibility}
E.~Mossel and M.~I. Ohannessian, ``On the impossibility of learning the missing
  mass,'' \emph{Entropy}, vol.~21, no.~1, p.~28, 2019.

\bibitem{cohen2021non}
S.~Cohen, T.~Routtenberg, and L.~Tong, ``Non-bayesian parametric missing-mass
  estimation,'' \emph{arXiv preprint arXiv:2101.04329}, 2021.

\bibitem{orlitsky2003always}
A.~Orlitsky, N.~P. Santhanam, and J.~Zhang, ``Always good turing:
  Asymptotically optimal probability estimation,'' \emph{Science}, vol. 302,
  no. 5644, pp. 427--431, 2003.

\bibitem{orlitsky2015competitive}
A.~Orlitsky and A.~T. Suresh, ``Competitive distribution estimation: Why is
  {G}ood-{T}uring good.'' in \emph{NIPS}, 2015, pp. 2143--2151.

\bibitem{cohen2017cardinality}
R.~Cohen, L.~Katzir, and A.~Yehezkel, ``Cardinality estimation meets
  {G}ood-{T}uring,'' \emph{Big data research}, vol.~9, pp. 1--8, 2017.

\bibitem{ayed2019good}
F.~Ayed, M.~Battiston, F.~Camerlenghi, S.~Favaro \emph{et~al.}, ``A
  {G}ood-{T}uring estimator for feature allocation models,'' \emph{Electronic
  Journal of Statistics}, vol.~13, no.~2, pp. 3775--3804, 2019.

\bibitem{gnedin2007notes}
A.~Gnedin, B.~Hansen, J.~Pitman \emph{et~al.}, ``Notes on the occupancy problem
  with infinitely many boxes: general asymptotics and power laws,''
  \emph{Probability surveys}, vol.~4, pp. 146--171, 2007.

\bibitem{barbour2009small}
A.~Barbour, A.~Gnedin \emph{et~al.}, ``Small counts in the infinite occupancy
  scheme,'' \emph{Electronic Journal of Probability}, vol.~14, pp. 365--384,
  2009.

\bibitem{blanchard2019accurate}
P.~Blanchard, D.~J. Higham, and N.~J. Higham, ``Accurate computation of the
  log-sum-exp and softmax functions,'' \emph{arXiv preprint arXiv:1909.03469},
  2019.

\bibitem{bu2018estimation}
Y.~Bu, S.~Zou, Y.~Liang, and V.~V. Veeravalli, ``Estimation of kl divergence:
  Optimal minimax rate,'' \emph{IEEE Transactions on Information Theory},
  vol.~64, no.~4, pp. 2648--2674, 2018.

\bibitem{boucheron2013concentration}
S.~Boucheron, G.~Lugosi, and P.~Massart, \emph{Concentration inequalities: A
  nonasymptotic theory of independence}.\hskip 1em plus 0.5em minus 0.4em\relax
  Oxford university press, 2013.

\bibitem{feng2013mean}
C.~Feng, H.~Wang, Y.~Han, Y.~Xia, and X.~M. Tu, ``The mean value theorem and
  taylor’s expansion in statistics,'' \emph{The American Statistician},
  vol.~67, no.~4, pp. 245--248, 2013.

\bibitem{boyd2004convex}
S.~Boyd, S.~P. Boyd, and L.~Vandenberghe, \emph{Convex optimization}.\hskip 1em
  plus 0.5em minus 0.4em\relax Cambridge university press, 2004.

\bibitem{biegler2010nonlinear}
\BIBentryALTinterwordspacing
L.~Biegler, \emph{Nonlinear Programming: Concepts, Algorithms, and Applications
  to Chemical Processes}, ser. MOS-SIAM Series on Optimization.\hskip 1em plus
  0.5em minus 0.4em\relax Society for Industrial and Applied Mathematics, 2010.
  [Online]. Available: \url{https://books.google.at/books?id=ZmIC7w9QnPEC}
\BIBentrySTDinterwordspacing

\bibitem{bazaraa2013nonlinear}
M.~S. Bazaraa, H.~D. Sherali, and C.~M. Shetty, \emph{Nonlinear programming:
  theory and algorithms}.\hskip 1em plus 0.5em minus 0.4em\relax John Wiley \&
  Sons, 2013.

\bibitem{lovric2007vector}
\BIBentryALTinterwordspacing
M.~Lovric, \emph{Vector Calculus}.\hskip 1em plus 0.5em minus 0.4em\relax
  Wiley, 2007. [Online]. Available:
  \url{https://books.google.at/books?id=hDdyDwAAQBAJ}
\BIBentrySTDinterwordspacing

\bibitem{2020SciPy-NMeth}
P.~Virtanen, R.~Gommers, T.~E. Oliphant, M.~Haberland, T.~Reddy, D.~Cournapeau,
  E.~Burovski, P.~Peterson, W.~Weckesser, J.~Bright, S.~J. {van der Walt},
  M.~Brett, J.~Wilson, K.~J. Millman, N.~Mayorov, A.~R.~J. Nelson, E.~Jones,
  R.~Kern, E.~Larson, C.~J. Carey, {\.I}.~Polat, Y.~Feng, E.~W. Moore,
  J.~{VanderPlas}, D.~Laxalde, J.~Perktold, R.~Cimrman, I.~Henriksen, E.~A.
  Quintero, C.~R. Harris, A.~M. Archibald, A.~H. Ribeiro, F.~Pedregosa, P.~{van
  Mulbregt}, and {SciPy 1.0 Contributors}, ``{{SciPy} 1.0: Fundamental
  Algorithms for Scientific Computing in Python},'' \emph{Nature Methods},
  vol.~17, pp. 261--272, 2020.

\bibitem{battiston2021consistent}
M.~Battiston, F.~Ayed, F.~Camerlenghi, and S.~Favaro, ``Consistent estimation
  of small masses in feature sampling,'' \emph{Journal of Machine Learning
  Research}, 2021.

\end{thebibliography}

\end{document}